\documentclass[twocolumn]{aastex631}

\newcommand\Msunperyear{M_\odot\,{\rm yr}^{-1}}

\received{April 22, 2025}
\accepted{July 9, 2025}

\submitjournal{ApJL}

\usepackage[normalem]{ulem}
\usepackage{threeparttable}
\usepackage{amsmath}

\usepackage{xcolor}
\DeclareRobustCommand{\erase}{\bgroup\markoverwith{\textcolor{red}{\rule[.5ex]{2pt}{0.4pt}}}\ULon}

\usepackage{float}

\shorttitle{Inferring Confined CSM Through Long-term SN Evolution}
\shortauthors{Matsuoka, Maeda, \& Chen}

\graphicspath{{./}{figure/}}

\begin{document}

\title{Inferring Dense Confined Circumstellar Medium around Supernova Progenitors\\via Long-term Hydrodynamical Evolution}

\author[0000-0002-6916-3559]{Tomoki Matsuoka}
\correspondingauthor{Tomoki Matsuoka}
\email{tmatsuoka@asiaa.sinica.edu.tw}
\affiliation{Institute of Astronomy and Astrophysics, Academia Sinica, No.1, Sec.4, Roosevelt Road, Taipei 106216, Taiwan}

\author[0000-0003-2611-7269]{Keiichi Maeda}
\affiliation{Department of Astronomy, Kyoto University, Kitashirakawa-oiwakecho, Sakyo-ku, Kyoto 606-8502, Japan}

\author[0000-0002-4848-5508]{Ke-Jung Chen}
\affiliation{Institute of Astronomy and Astrophysics, Academia Sinica, No.1, Sec.4, Roosevelt Road, Taipei 106216, Taiwan}

\begin{abstract}
Circumstellar interaction of supernova (SN) ejecta is an essential process in its evolution and observations of SNe have found the signature of circumstellar interaction both in the early and late evolutionary phase of SNe. In this {\it Letter}, we show that if the SN forward shock plunges into tenuous stellar wind from dense circumstellar medium (CSM) in the vicinity of the progenitor (i.e., confined CSM), the subsequent time evolutions of the SN-CSM interaction system deviates from the prediction of self-similar solution. In this case, after all of the confined CSM is swept up by the SN forward shock (roughly $10\,{\rm days}$ after the explosion), the propagation of the shocked shell will be driven by the freely expanding ram pressure of the confined CSM component, instead of the SN ejecta. Meanwhile, the forward shock decelerates faster than the prediction of thin-shell approximation once the confined CSM component reaches homologous expansion. This lasts until the reverse shock in the confined CSM component reaches the head of the SN ejecta, leading to the restoration of the system into the evolutionary model without confined CSM, where the SN ejecta drives the expansion of the system. We also show that this peculiar evolution will be reflected in observational signatures originating from SN-CSM interaction, taking rapid decline and rebrightening of radio emission as examples. Our results shed light on the importance of taking into account the effect of initial SN-CSM interaction even when we focus on observational properties of SNe a few years after the explosion.
\end{abstract}


\section{Introduction}\label{sec:introduction}
Circumstellar interaction in supernovae (SNe) is a fundamental process that largely characterizes the evolution of the SN itself; it largely moderates the expansion of the SN ejecta irrespective of the mass of the circumstellar medium (CSM) relative to the SN ejecta itself \citep{1982ApJ...258..790C}.
CSM interaction produces unique observational signatures emitted in a wide range of wavelength; enhancement of optical luminosity \citep[e.g.,][]{2013MNRAS.435.1520M,2017MNRAS.469L.108M}, emergence of narrow lines in the spectra \citep[e.g.,][]{1997ARAA..35..309F,2015MNRAS.449.4304D}, and non-thermal emissions \citep[e.g.,][]{2017hsn..book..875C}. These radiative properties are essential for understanding the nature of CSM interaction which is also a vital tracer of mass-loss activities of SN progenitors.

Recent rapid follow-up photometric and spectroscopic observations have indicated that a large fraction of type II SN progenitors experience enormous mass-loss activity ($\dot{M}\gtrsim 10^{-3}\,M_\odot{\rm yr}^{-1}$) immediately before the explosion ($1-100$~years). This picture is evidenced by detections of highly-ionized narrow flash features in optical spectra and early excess in light curves of optical/infrared bands within a few days after the explosion, as proven in the representative cases of SN\,2013fs \citep{2017NatPh..13..510Y}, 2021gmj \citep[][]{2024MNRAS.528.4209M}, 2023ixf \citep[e.g.,][]{2023ApJ...954L..42J,2023ApJ...955L...8H,2023ApJ...956L...5B,2024ApJ...975..132S}, and 2024ggi \citep{2024ApJ...972..177J,2024ApJ...972L..15S}. 
These pieces of observational evidence imply the presence of CSM distributed in the vicinity of the progenitor, which is believed to be by orders of magnitude denser than predictions from the canonical stellar wind theory.
This casts a new picture of massive stars' evolution where they may experience a drastic increase in the mass-loss rate immediately before the explosion. A similar trend has been found in the case of a stripped-envelope SN 2020oi, in which the temporal variation of the mass-loss rate by a factor in a Wolf-Rayet star has been inferred from radio observations \citep{2021ApJ...918...34M}.

Investigations of SN-CSM interaction are not limited to the initial phase but also expanded in the late phase of SN evolution. Previous studies have found strong CSM interaction signatures in some SNe appearing a few years after the explosion through excess of optical emission \citep{2020ApJ...900...11W, 2022ApJ...936..111K} and {significant} rebrightening of radio emission \citep{2017ApJ...835..140M, 2023ApJ...954L..45M, 2023ApJ...945L...3M,2024MNRAS.534.3853R,2025arXiv250201740S}. These evidences highlight a possibility that massive stars' eruptive activity may be also realized several thousands of years before the explosion. The possible physical processes producing such an eruptive activity include excitation of gravity wave in the stellar core \citep{2012MNRAS.423L..92Q,2014ApJ...780...96S,2017MNRAS.470.1642F}, explosive shell burning instabilities \citep{2011ApJ...733...78A, 2014ApJ...785...82S}, and binary interaction \citep{2012ApJ...752L...2C,2024ApJ...963..105M,2024A&A...685A..58E,2024arXiv241209893E}. {Not limited to radio rebrightening, the modulation of observed radio luminosity detected during its decay phase has been reported ever and interpreted as a possible density variation of the CSM \citep{2004MNRAS.349.1093R,2006ApJ...651.1005S}.}

In general, when we attempt to constrain the CSM interaction properties, it is often assumed that observational signatures obtained in the later phase are dominated by the SN-CSM interaction configuration at the given time. This is equivalent to the idea that the system has forgotten the history of the past SN-CSM interaction. However, based on long-term, one-dimensional hydrodynamics simulations of SN-CSM interaction, we show that the late-phase SN evolution can be actually affected if the SN progenitor possesses the so-called confined CSM; the dense CSM in the vicinity of the progenitor. Initially the forward shock is formed in the confined CSM, but after it plunges into the outside tenuous CSM, the system rapidly expands and the spatial distribution of the shocked region is stretched.
Once the confined CSM component reaches homologous expansion profile, the forward shock starts faster deceleration than the prediction of self-similar solution. {Afterwards}, the reverse shock propagating through the confined CSM component reaches the head of the SN ejecta, causing the increase in the forward shock velocity by a factor, {and} restoring the evolutionary model without confined CSM.
We also show that these peculiar evolutionary properties would be reflected in observational signatures originating from CSM interaction, taking radio emission as an example. Our demonstration sheds light on the importance of taking into account the past history of SN-CSM interaction dynamics even if we focus on the late-phase properties of SN-CSM interaction.

\begin{table}[t]
    \centering
    \caption{CSM models examined in this {\it Letter}.}
    \begin{tabular}{c|cc}
    \hline
    \hline
    Model & $\dot{M} (M_\odot\,{\rm yr}^{-1})$ & $M_{\rm CSM}\,(M_\odot)$ \\
    \hline
     High density CSM & $10^{-3}$ & $3\times 10^{-2}$\\
     Intermediate density CSM & $10^{-4}$ & $3\times 10^{-3}$\\
     Low density CSM & $10^{-5}$ & $3\times 10^{-4}$\\
     \hline
     No confined CSM & $10^{-6}$ & $(3\times 10^{-5})$\\
    \hline
    \hline
    \end{tabular}
    \label{tab:CSMmodels}
\end{table}

This {\it Letter} is organized as follows. Section\,\ref{sec:method} introduces the setup and method of our numerical simulations. Section\,\ref{sec:results} showcases the results of our numerical simulations with the responsible physical processes for the peculiar evolution described. We also calculate the expected evolution of the observed radio luminosity in our models, in order to practically show that the effects shown in our simulations can clearly appear in observational signatures. Finally, Section\,\ref{sec:summary} summarizes the contents of our {\it Letter} with some discussion described.

\section{Setup and Method}\label{sec:method}

\begin{figure*}
    \centering
    \includegraphics[width=0.45\linewidth]{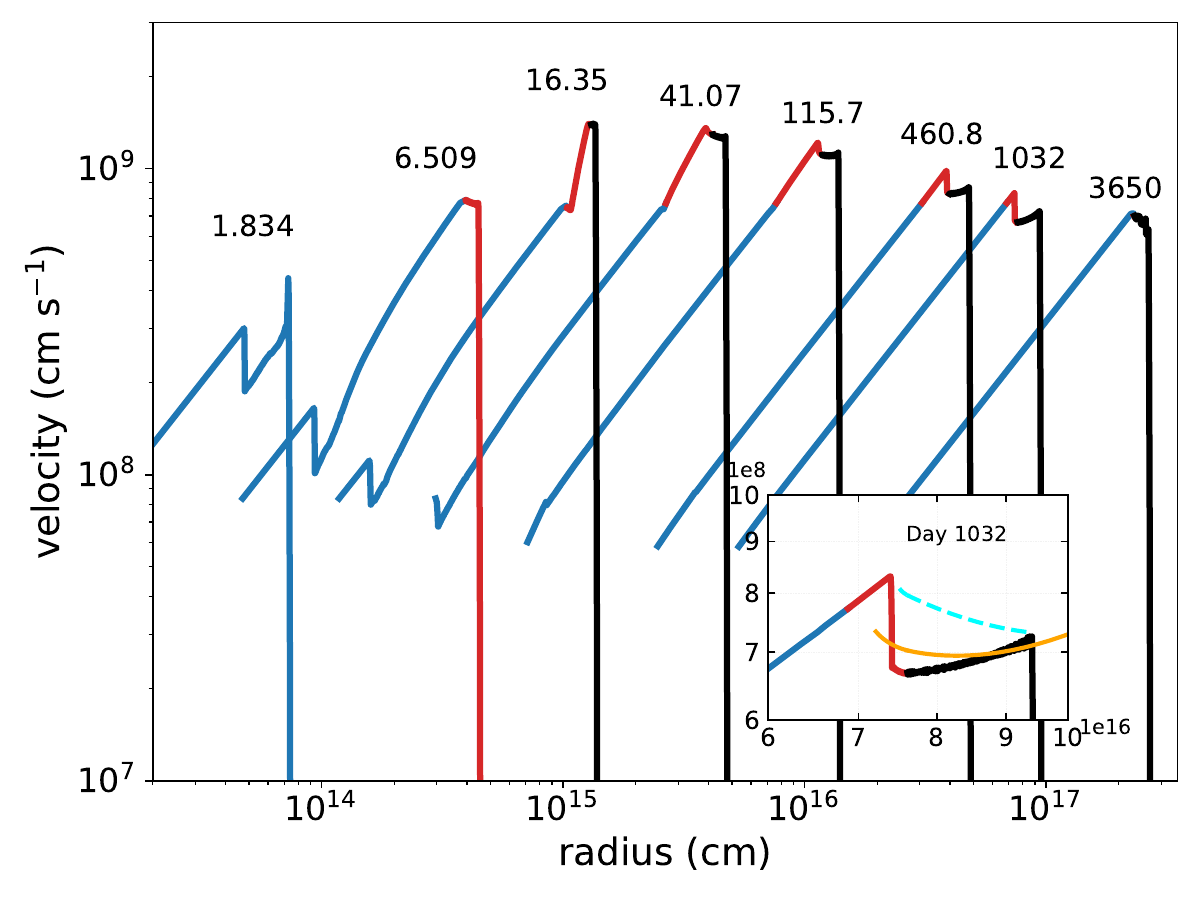}
    \includegraphics[width=0.45\linewidth]{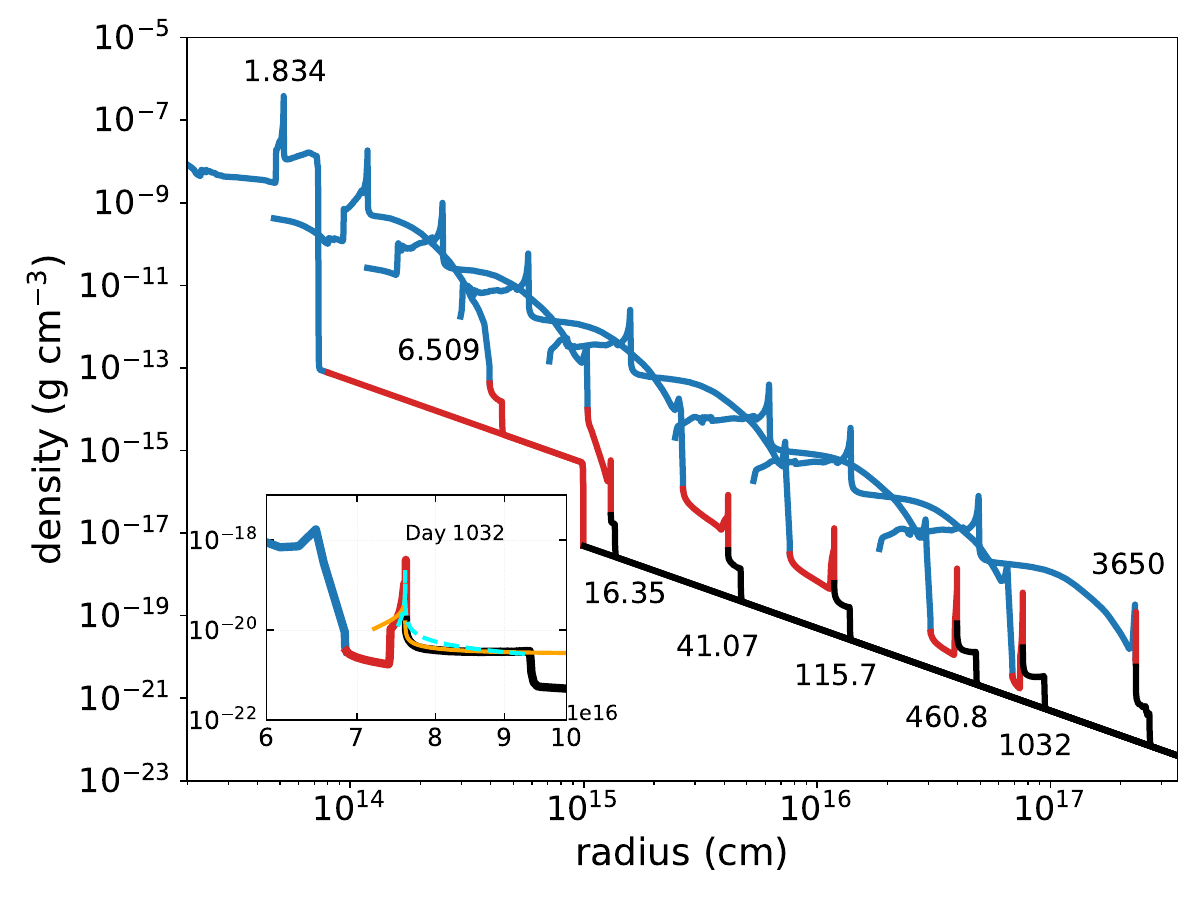}
    \includegraphics[width=0.45\linewidth]{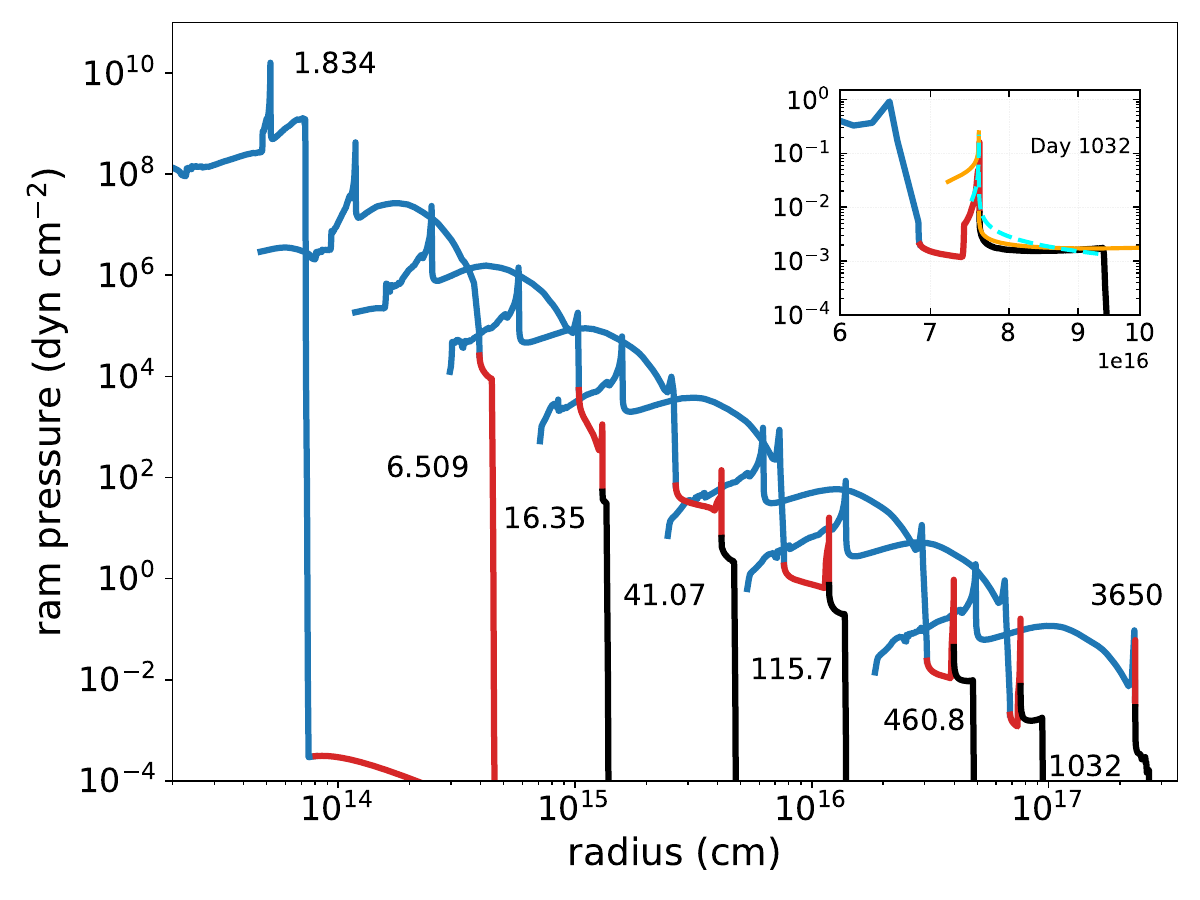}
    \includegraphics[width=0.45\linewidth]{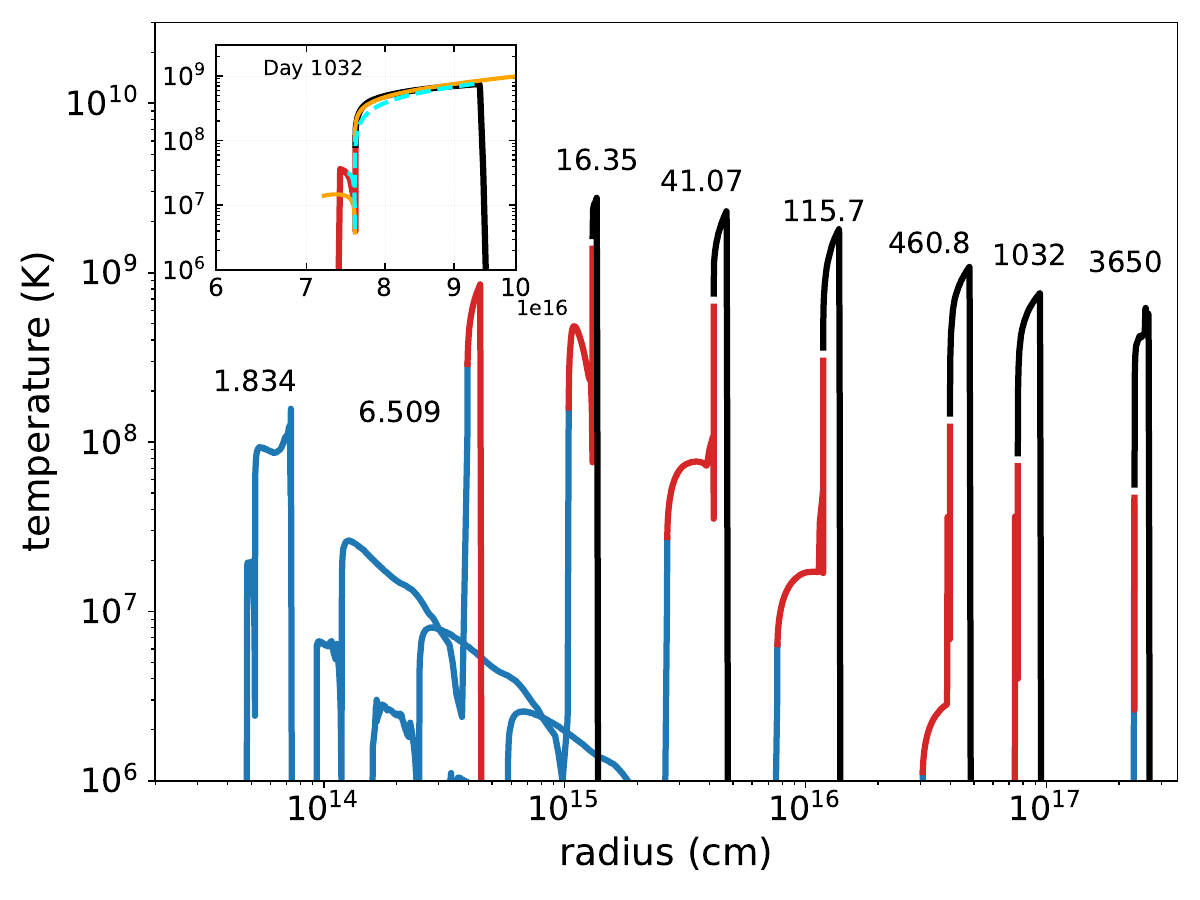}
    \caption{Time evolution of velocity (upper left), density (upper right), ram pressure (lower left), and temperature (lower right) profiles in the SN-CSM interaction model possessing confined CSM with the mass-loss rate of $\dot{M}=10^{-4}\,\Msunperyear$. The SN ages are noted alongside each profile in the unit of day. Blue, red, and black lines indicate that the gas component is the SN ejecta, confined CSM, and tenuous wind CSM, respectively. The snapshots at the SN age of {1032}~days are zoomed in {and compared with Chevalier's self-similar solutions with $n=5.5$ (solid orange) and $n=12$ (cyan dashed).}}
    \label{fig:hydrodynamics_profile}
\end{figure*}

We conduct nonradiative hydrodynamics of the expansion of SN ejecta colliding with various kinds of CSM structures.
In this study, we focus on type II SNe for the purpose of demonstration, but we expect our results can be applied to the other SN types once the appropriate conditions and setups have been addressed. We prepare a type II SN progenitor model originated from the zero-age main sequence (ZAMS) mass of $15\,M_\odot$, constructed by the stellar evolutionary simulation code, MESA \citep{2011ApJS..192....3P, 2013ApJS..208....4P, 2015ApJS..220...15P, 2019ApJS..243...10P}. This progenitor model takes stellar wind mass loss into account, and thus the stellar mass at the moment of the core collapse is $\simeq 12\,M_\odot$. Then we attach a range of CSM structures to the progenitor. We examine four models for the CSM given as follows. As a basic component of the CSM, we consider stellar wind from the progenitor. Since the progenitor is considered to be an inflated red supergiant with low effective temperature (typically $\sim4000{\rm K}$), we suppose the wind CSM would be characterized by the mass-loss rate of $\dot{M}=10^{-6}\,\Msunperyear$ with the wind velocity of $v_w = 10\,{\rm km\,s}^{-1}$ \citep{1988A&AS...72..259D}.
Then we place the additional CSM component within the specific lengthscale of $R_{\rm CSM}$, which serves as a confined CSM. Here we increase the corresponding mass-loss rate to $\dot{M}=10^{-5},10^{-4}$, and $10^{-3}\,\Msunperyear$ within the lengthscale of $R_{\rm CSM} = 10^{15}\,{\rm cm}$, while the velocity of this component is fixed. The corresponding mass of the confined CSM ($M_{\rm CSM}$) amounts to $3\times 10^{-4},3\times 10^{-3}$, and $3\times 10^{-2}\,M_\odot$, respectively.
Now we have three CSM models with the confined CSM accompanied, and we further examine the reference model in which there is no density enhancement in the CSM. In total, we examine the hydrodynamical simulations with four CSM models. Table\,\ref{tab:CSMmodels} summarizes the four CSM models examined in this {\it Letter}.

We employ the open code SNEC \citep{2015ApJ...814...63M} to conduct numerical simulations of nonradiative hydrodynamics of the interaction between SN ejecta and CSM, which is designed on the basis of the Lagrangian description for fluid dynamics. We excise the inner region containing the mass budget of $1.4\,M_\odot$ to mimic the formation of a proto-neutron star. Then we inject {thermal energy with parameters \tt{bomb\_mass\_spread=0.3d0} and \tt{final\_energy=1.0d51}}, and follow the time evolution of the system, particularly focusing on the temporal evolutions of hydrodynamical structure and forward shock. We terminate our simulations at the SN age of 10~years.

{It is recognized that radiative cooling would play an important role in the formation of the thin shell in the shocked region, producing a range of thermal and nonthermal emissions \citep[e.g.,][]{1982ApJ...259..302C}. The formation of the cooling shell would evidently deviate the hydrodynamical profiles from the self-similar solution developed by \cite{1982ApJ...258..790C} because it does not include the effect of radiative cooling. We discuss the possible influence of radiative cooling on our results in Section~\ref{sec:summary}.}

\section{Results}\label{sec:results}
{\subsection{Hydrodynamical profiles}}\label{subsec:hydro_profile}
Figure\,\ref{fig:hydrodynamics_profile} shows the time evolution of profiles of velocity, density, ram pressure, and temperature of SN ejecta interacting with the CSM involving confined CSM with the mass-loss rate of $10^{-4}\,\Msunperyear$. The first snapshot illustrates the moment of the shock breakout from the progenitor, showing that the ejecta is about to start colliding with CSM ($t=1.83\,{\rm days}$).
The interaction between the SN ejecta and confined CSM (shown at $t=6.51\,{\rm days}$) lasts until the forward shock reaches the outer edge of the confined CSM (${R_{\rm FS}} \simeq R_{\rm CSM}$). The hydrodynamical profile is still in the way of reaching the {nonradiative} self-similar solution profile.

\begin{figure}
    \centering
    \includegraphics[width=\linewidth]{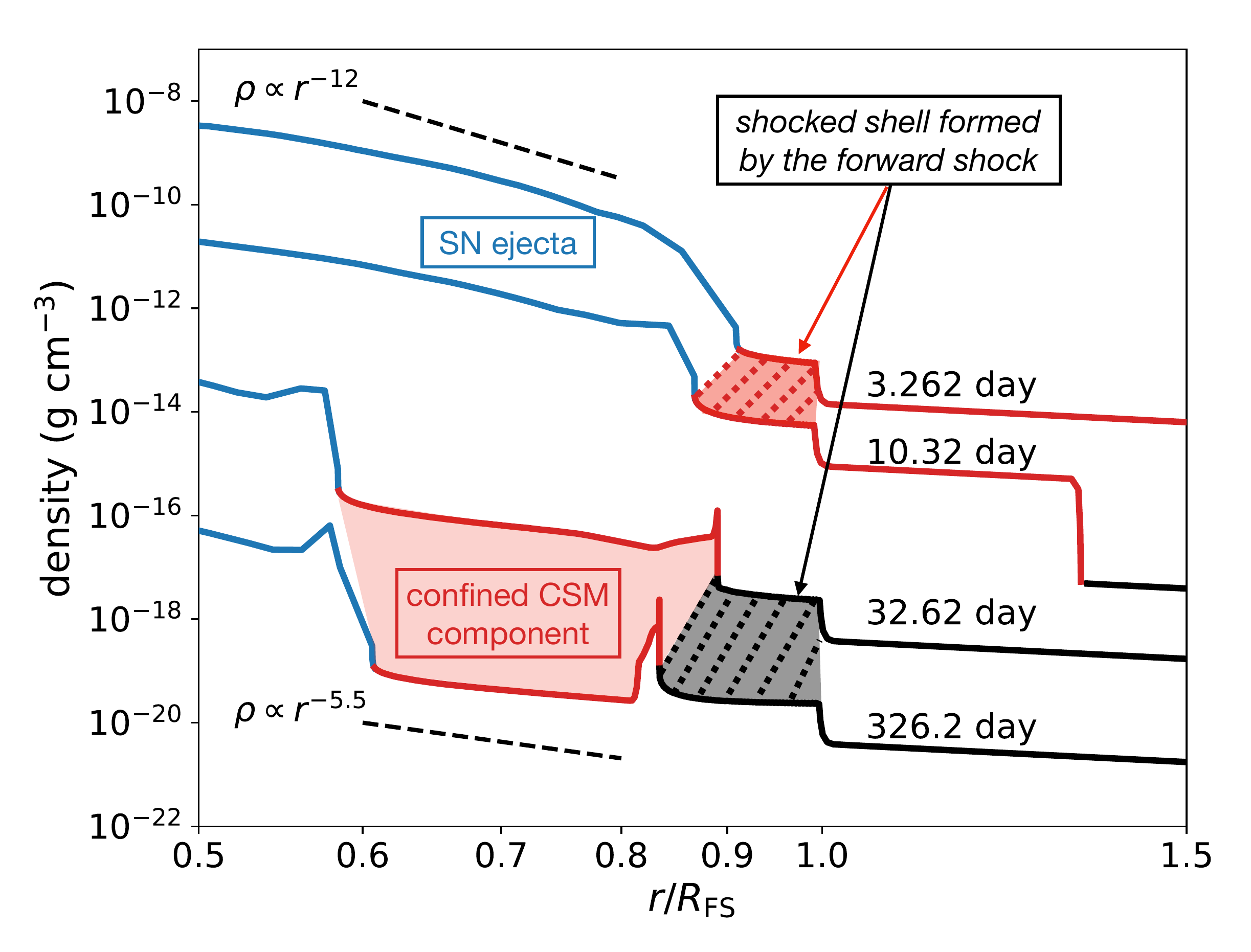}
    \includegraphics[width=\linewidth]{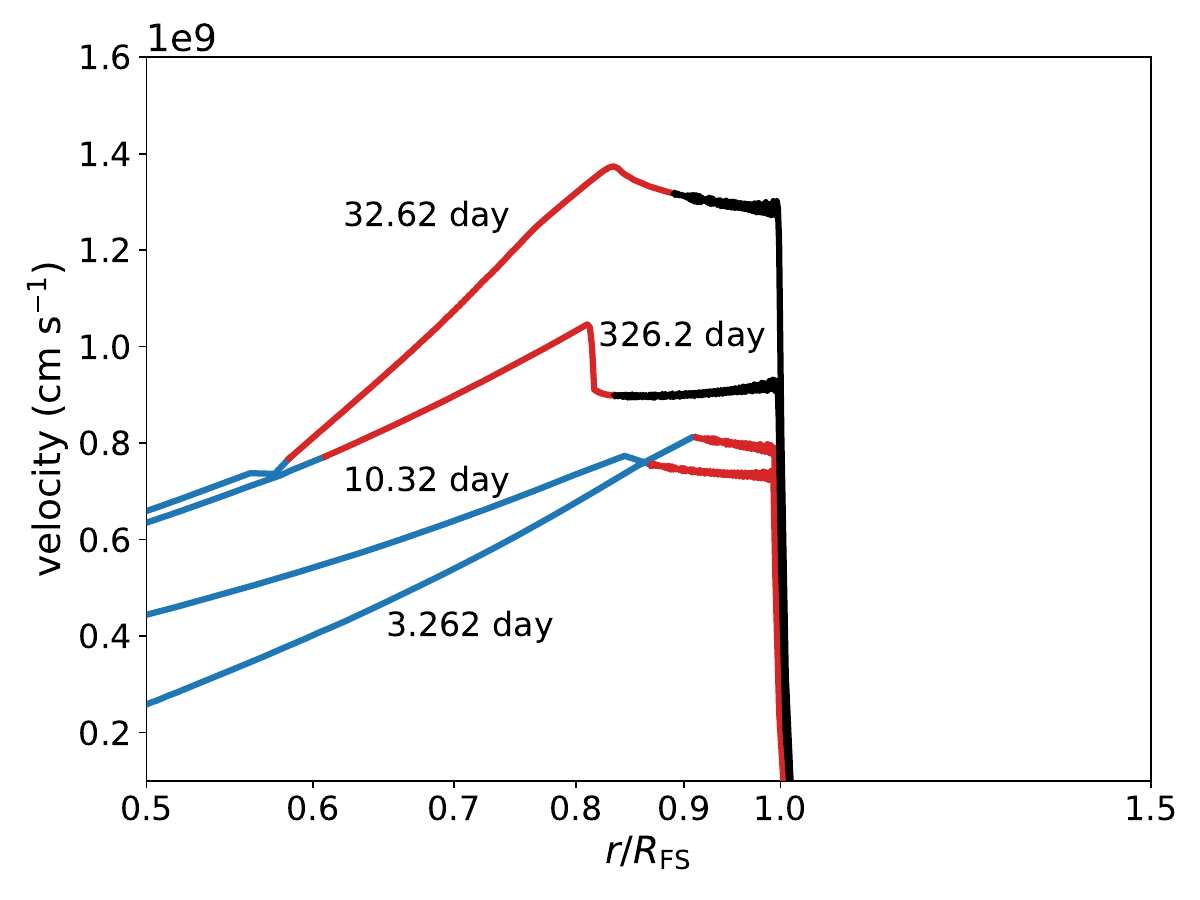}
    \caption{Top: Time evolution of the density structure with the radius in the abscissa normalized by the forward shock radius at given times. The color usage is the same as Figure\,\ref{fig:hydrodynamics_profile}. Before $t\lesssim 10\,{\rm days}$ the forward shock compresses the confined CSM component and the heated confined CSM component is indicated by the red shaded region. After $t\gtrsim10\,{\rm days}$ the newly formed forward shock heats up the tenuous wind, which is shaded in the black region.
    The black dashed lines denote the reference density profiles of the ejecta with $n=12$ and $n=5.5$. Bottom: Time evolution of the velocity structure with the radius normalized by the forward shock radius. Note that after $t\gtrsim 10\,{\rm days}$ the maximum velocity takes in the confined CSM component, indicating that it is a driving source for the expansion of the shocked shell.}
    \label{fig:flattening}
\end{figure}

Once the forward shock completes sweeping up the confined CSM, the system starts rapid expansion and energy redistribution is taking place in the confined CSM component. A new forward shock is formed, which propagates through the tenuous wind component residing outside of the confined CSM as shown in the profiles of $t=16.4,41.0\,{\rm days}$ in Figure\,\ref{fig:hydrodynamics_profile}. As a result, there are two components of the CSM that have experienced shock compression; the confined CSM component inside and the tenuous wind component outside.
Note that the shocked shell formed by the forward shock now consists of the tenuous wind and is expanding due to the ram pressure from the confined CSM component (see {also} Figure\,\ref{fig:flattening}). 
{This is distinct from the standard assumption of the self-similar solution} where the expansion is driven by the ram pressure of adiabatically {expanding} ejecta.
The confined CSM component reaches homologous expansion profile at $t\simeq115.7\,{\rm days}$, as found in the velocity profile where $v(r)\propto r$ is almost reconstructed. This can be confirmed by the fact that the density profile of the confined CSM component is becoming flatter between $16.4\,{\rm days}\lesssim t\lesssim41.0\,{\rm days}$, while it is almost unchanged in $41.0\,{\rm days}\lesssim t\lesssim115\,{\rm days}$.

\begin{figure}
    \centering
    \includegraphics[width=\linewidth]{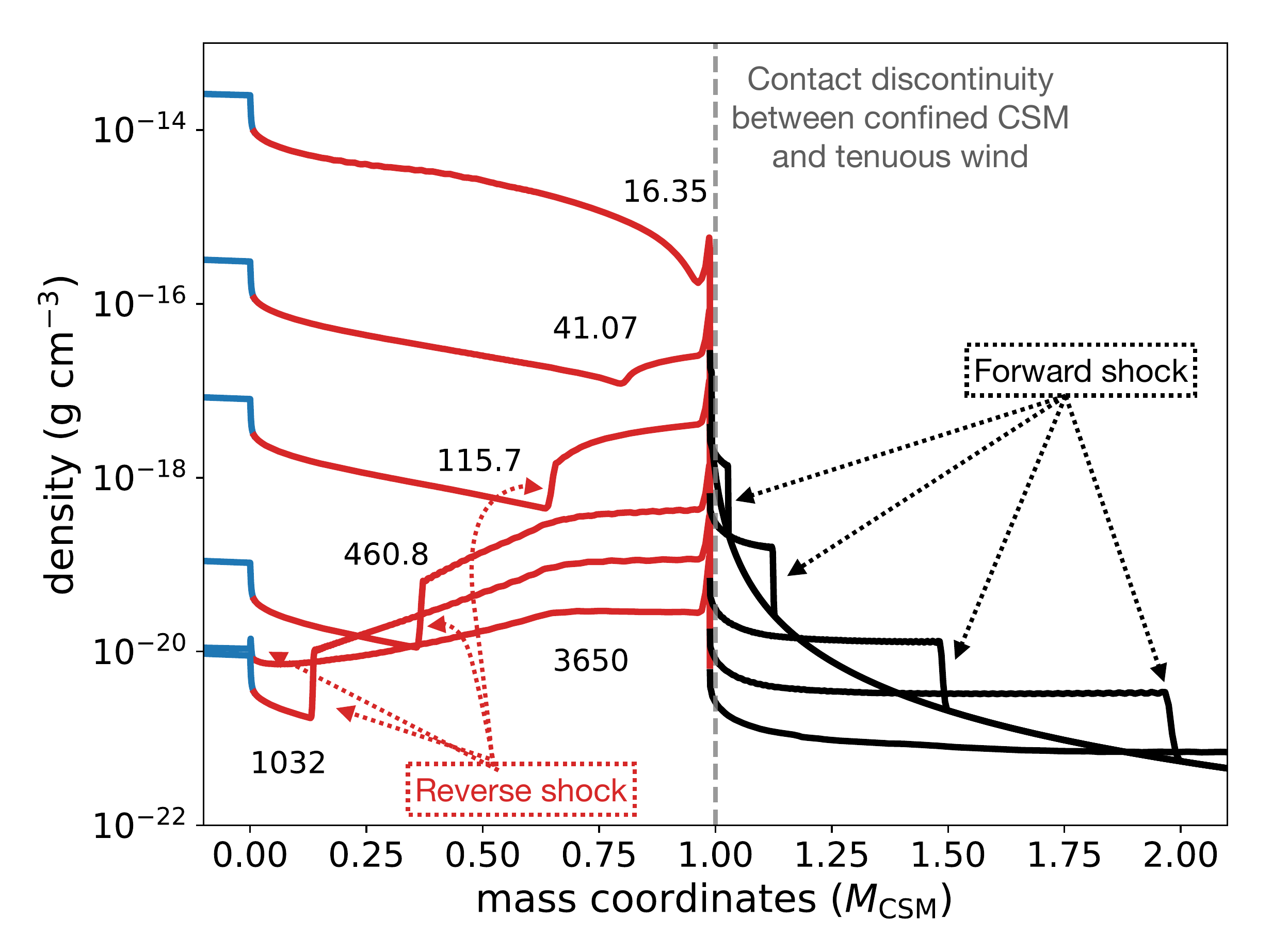}
    \caption{Time evolution of the density structure in the model with $\dot{M}=10^{-4}\,M_\odot{\rm yr}^{-1}$, but the x-axis is plotted by the mass coordinates {$(m)$} with the zero point set to the contact discontinuity between the SN ejecta and the confined CSM. The color usage is the same as Figure\,\ref{fig:hydrodynamics_profile}. $m=M_{\rm CSM}$ in this mass coordinates corresponds to the contact discontinuity between two types of CSM, while the discontinuities seen in the confined CSM and tenuous wind are the reverse and forward shock, respectively.}
    \label{fig:density_mass}
\end{figure}

As a result of the energy redistribution, the confined CSM component reaches homologous expansion and there is a reverse shock developing in the confined CSM component.
The reverse shock is associated with the forward shock propagating in the outside tenuous wind, and we can clearly find out the reverse shock in the zoomed panel of the hydrodynamical profile at $t={1032}\,{\rm days}$ in Figure\,\ref{fig:hydrodynamics_profile}. The expansion of the system is driven by the ram pressure of this homologously expanding confined CSM component ($t=460.8,\,1032\,{\rm days}$).
It is also illustrated in Figure\,\ref{fig:density_mass}, where density profile evolution is plotted as a function of the mass coordinates, and we can see that the reverse shock is propagating through the confined CSM inward.
Once this reverse shock reaches the head of the SN ejecta (zero point of the mass coordinates in Figure\,\ref{fig:density_mass}), the confined CSM component almost assimilates with the SN ejecta, and the driving source for the expansion changes over the SN ejecta ($t=3650\,{\rm days}$).

Figure\,\ref{fig:flattening} shows the density structure of the interacting region, but the x-axis is normalized by the radius of the forward shock at given times.
Before the forward shock arrives at the edge of the confined CSM, the expansion of the shocked shell (the region indicated by the dashed lines) is driven by the ram pressure of the SN ejecta ($t=3.262,10.32\,{\rm days}$). The temporal evolution of the shocked shell formed by the forward shock is considered to depend on the density gradient of the component inside this region; in this case, the gradient of the outer layer of the SN ejecta would be important. \citet[][]{1999ApJ...510..379M} proposed the feasible values of $n$ according to the progenitor, where $\rho_{\rm ej} \propto r^{-n}$ depending on the progenitor star. {Particularly, $n\simeq12$ has been employed for an explosion of a red supergiant \citep[e.g.,][]{2006ApJ...641.1029C} and our simulation is likely to be consistent with this prediction, as for this time range.} The situation changes once the newly generated forward shock starts propagating through the outer tenuous wind. Now the main component of the shocked shell is the tenuous wind CSM ($t=32.62, 326.2\,{\rm days}$), {and its expansion would be driven not by the SN ejecta, but by the confined CSM component that has experienced shock compression before. This is confirmed by the fact that the confined CSM component takes a maximum velocity in the profile.}


{It is also seen from Figure\,\ref{fig:flattening} that the confined CSM component has a density slope of $n\simeq 5.5$, which is flatter than the SN ejecta with $n\simeq 12$. However, we mention that the hydrodynamical profile seen in the shocked shell would be different from the nonradiative self-similar solution described in \citet{1982ApJ...258..790C}. The zoomed panels in Figure\,\ref{fig:hydrodynamics_profile} show the hydrodynamical profiles compared with the self-similar solution with $n=5.5, s=2$ and $n=12, s=2$, plotted by solid orange and cyan dashed lines, respectively. We find that the hydrodynamical distribution between the forward shock and contact discontinuity matches better to the solution of $n=5.5$. On the other hand, the profile between the reverse shock and contact discontinuity is roughly consistent with the self-similar solution with $n=12$. It is also seen that the geometrical thickness of the region between the forward shock and contact discontinuity is also compatible with that computed from the self-similar solution with $n=12$.
One of the reasons may be that the interacting region still possesses the information of the past CSM interaction. As shown in \citet{1982ApJ...258..790C}, the ratio of the masses of the materials swept by the reverse shock and forward shock (noted as ${M_2/M_1}$ in \citealp{1982ApJ...258..790C}) decreases as the ejecta slope becomes shallow. We can see from Figure\,\ref{fig:density_mass} that after the forward shock plunges into the tenuous wind, this mass ratio decreases as time goes by. This implies the possibility that we are observing the moment when the interacting region is in the transition phase from SN ejecta-driven expansion to confined CSM component-driven expansion.}

{\subsection{Forward shock velocity evolution}}\label{subsec:FS}

\begin{figure}
    \centering
    \includegraphics[width=\linewidth]{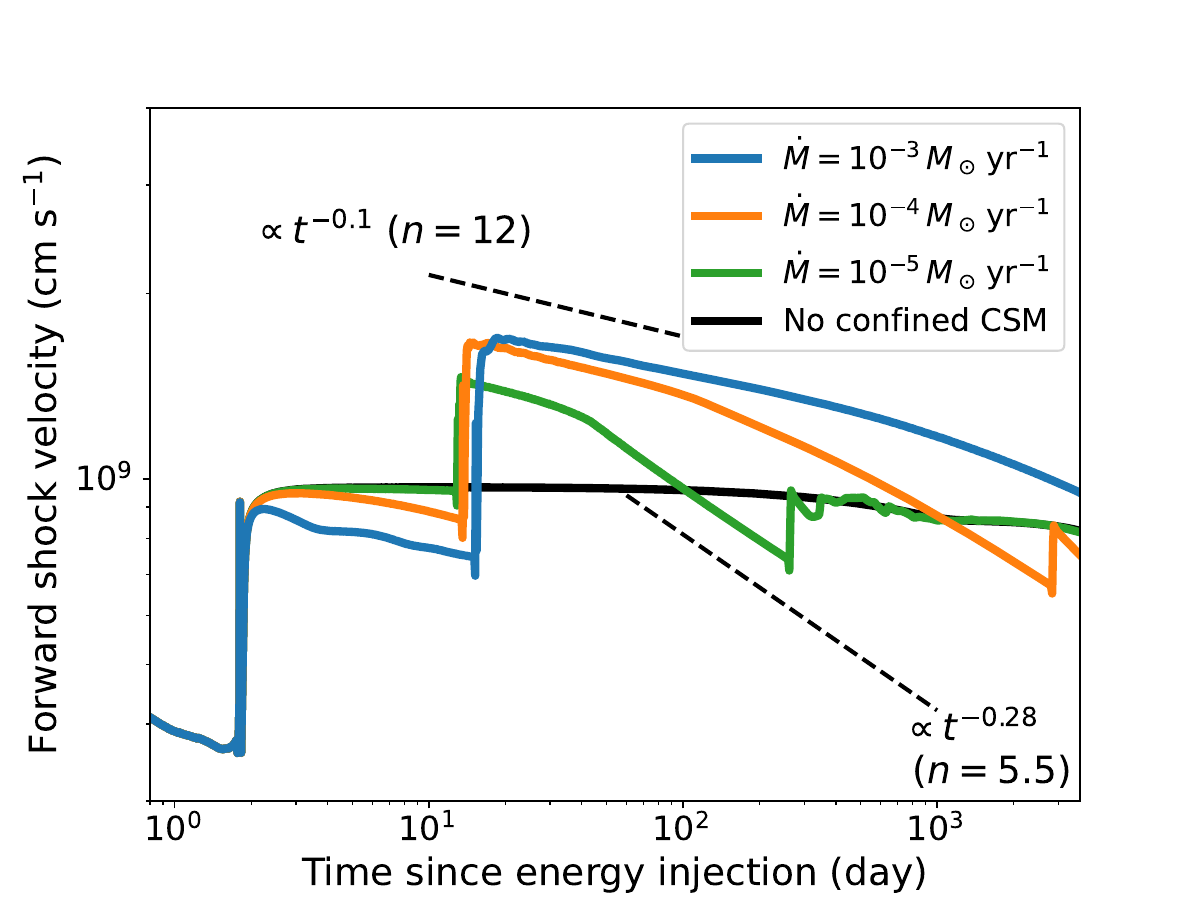}
    \caption{Time evolution of the forward shock velocity in our models. The dashed black lines denote the time dependence of $\propto t^{-0.1}$ (corresponding to $n=12$) and $\propto t^{-0.28}$ ($n=5.5$) for the reference.}
    \label{fig:Vshock}
\end{figure}

{The peculiar density structural evolution discussed in Section~\ref{subsec:hydro_profile}} leaves an imprint on the velocity evolution of the forward shock, which is illustrated in Figure\,\ref{fig:Vshock} as a function of time since the explosion energy injection. Here we mention the power-law index of the temporal evolution of the forward shock, which is computed as follows:
\begin{eqnarray}
m-1 = \frac{s-3}{n-s}
\end{eqnarray}
where $m=(n-3)/(n-s)$ is the deceleration parameter representing the power-law index of the temporal evolution of the forward shock radius, and $s=2$ is the density slope of the CSM. This parameter serves as a useful indicator in the following discussion.

We briefly explain the temporal evolution of the shock velocity observed immediately after the shock breakout from the stellar surface at $t\sim2\,{\rm days}$. In the models of $\dot{M}=10^{-3}\,M_\odot{\rm yr}^{-1}$ and $10^{-4}\,M_\odot{\rm yr}^{-1}$, both of which have relatively massive confined CSM, the density at the head of the SN ejecta is already lower than the density of the CSM swept up by the shock. This situation corresponds to ejecta-dominated phase leading to the solution with the slowly decelerating shock velocity \citep[][]{1982ApJ...258..790C,1999ApJS..120..299T}. The temporal evolutions in the numerical models are consistent with the prediction from the self-similar solution (${V_{\rm FS}}\propto t^{-0.1}$, see Figure\,\ref{fig:Vshock}), except the difference in the normalization of the velocity originating from the density scale of the confined CSM. 
On the other hand, in the models of $10^{-5}\,M_\odot{\rm yr}^{-1}$ and without confined CSM, the shock velocity remains roughly constant within $t\lesssim 10\,{\rm days}$. This is caused by the artificial setup that our progenitor model is not resolved {in the atmosphere} where the density steeply drops down; in this case, the density at the head of the ejecta is higher than {that} of the CSM swept up by the shock immediately after the shock breakout from the progenitor. This situation corresponds to free-expansion phase where the SN shock does not experience deceleration. In fact, radiative cooling at the moment of shock breakout from the progenitor surface would carry out a fraction of energy from the head of the ejecta, making the forward shock decelerate \citep{1999ApJ...510..379M,2013ApJ...762...14M}. This phenomenon is apt to be realized in the system where the CSM density is low and the system is optically thin to the radiation. In any case, the temporal evolutions of the forward shock in these models are in line with our classical understanding.

The important phenomenon can be observed after the forward shock plunges into the tenuous wind.
At this moment, the velocity of the forward shock suddenly increases by a factor ($t\sim 10\,{\rm days}$) because of the discontinuous drop in the CSM density \citep[see also][]{2019ApJ...885...41M}. The afterward evolution differs among the models. For example in the model with the mass-loss rate of $\dot{M}=10^{-5}\,\Msunperyear$, the forward shock starts the faster deceleration at $t\sim40\,{\rm days}$ since the energy injection. During this fast deceleration, the forward shock velocity is roughly proportional to $\propto t^{-0.28}$, corresponding to that derived from the self-similar solution with $n=5.5$ and $s=2$.
Further, the forward shock velocity increases by $\sim30\%$, as found at $t\sim 270\,{\rm days}$.
After the restoration process of the system accompanied by the fluctuation of ${V_{\rm FS}}$, the evolution of the forward shock becomes very similar to the model without the confined CSM. Similar fast deceleration can be found in the cases of $\dot{M}=10^{-3}\,M_\odot{\rm yr}^{-1}$ and $10^{-4}\,M_\odot{\rm yr}^{-1}$. While the convergence to the {model without confined CSM} is observed in the model of $10^{-4}\,M_\odot{\rm yr}^{-1}$ at $t\simeq2900\,{\rm days}$, we expect the same phenomenon will be realized even in the model with $10^{-3}\,M_\odot{\rm yr}^{-1}$ after $t=3650\,{\rm days}$.

{The fast deceleration of the forward shock starts at around $t\simeq300, 100, 40\,{\rm days}$ in the models of $\dot{M}=10^{-3}, 10^{-4}, 10^{-5}\Msunperyear$, respectively. We find that notably, these timings are likely to be synchronized with the moment when the confined CSM component reaches the homologous expansion profile with the velocity distribution following $v(r)\propto r$. We suspect that the fast deceleration of the forward shock is associated with the temporal variation of the hydrodynamical profile of the confined CSM component. It is also worth noting that the more massive the CSM mass is, the more delayed the deceleration timing becomes. The initiation timescale of the fast deceleration is likely to be proportional to $\propto M_{\rm CSM}^{1/2}$ at most within a factor.}

The increase in the forward shock velocity is associated with the arrival of the reverse shock in the confined CSM component at the head of the SN ejecta (see Figure\,\ref{fig:density_mass}). The reach of the reverse shock indicates that the confined CSM component is already assimilated with the SN ejecta and now it is the SN ejecta that serves as a ram pressure source for driving the shocked shell expansion. As a result, the characteristics of the confined CSM component vanish and the subsequent evolution converges to the model with only the tenuous wind. The timing of the increase in the forward shock velocity would be when the mass budget of the tenuous wind component heated by the forward shock has {roughly} amounted up to the total confined CSM mass.

\begin{figure}
    \centering
    \includegraphics[width=\linewidth]{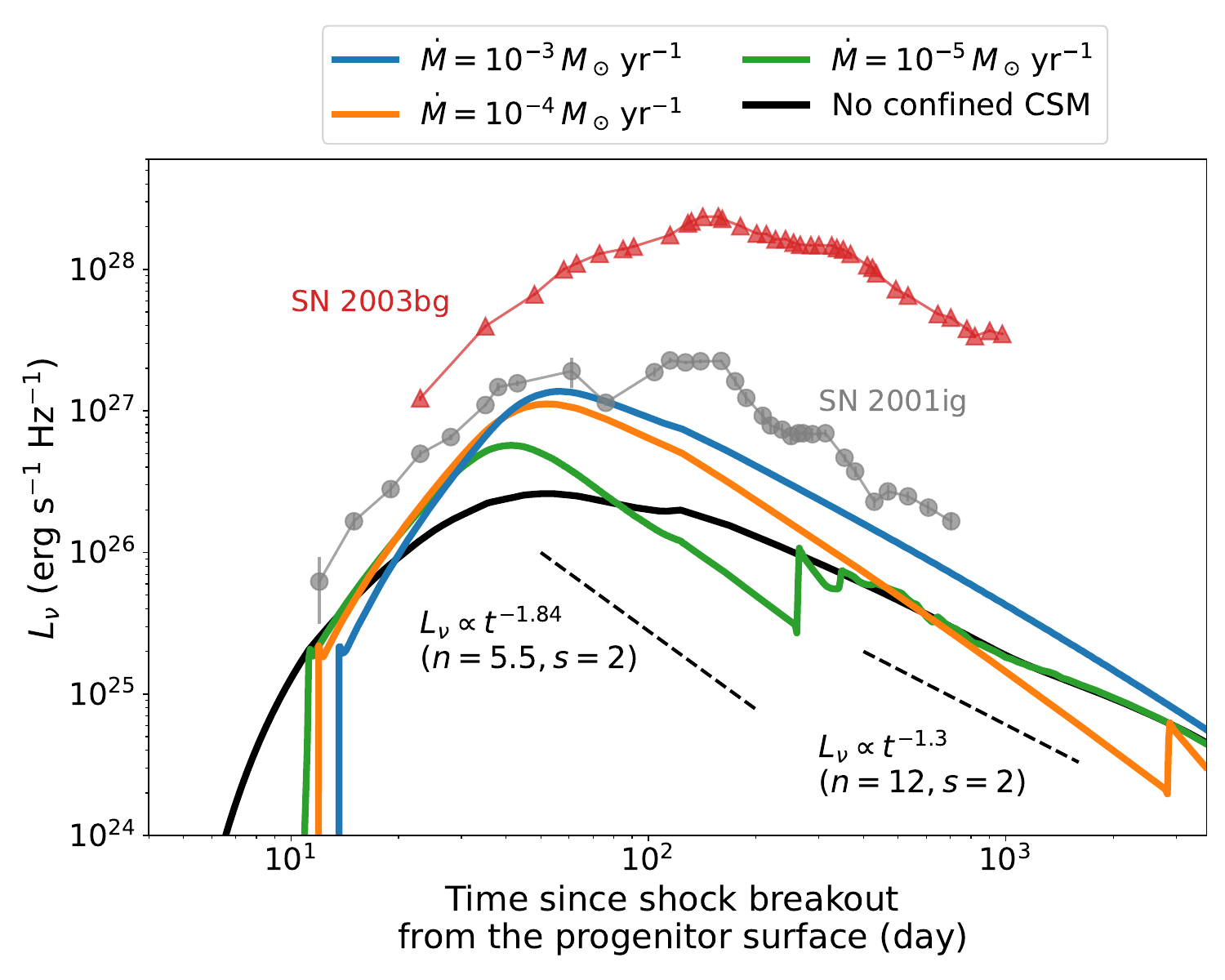}
    \caption{Radio light curves constructed from our models. The black lines denote the temporal slope of optically thin emission in an adiabatic cooling phase. Note that the x-axis is the time since the shock breakout from the progenitor. {Gray and red points are the observational data of SN~2001ig \citep{2004MNRAS.349.1093R} and SN~2003bg \citep{2006ApJ...651.1005S} at $4.8\,$GHz, respectively.}}
    \label{fig:Lnu}
\end{figure}

{\subsection{Radio emission as an observational counterpart}}\label{subsec:radio}

Finally, we show an example of how {the} peculiar forward shock evolution affects the observational signature. Figrue\,\ref{fig:Lnu} shows the radio {light curves for each model at the frequency of 5\,GHz, which is one of the typical frequencies adopted in radio SN observations \citep{2021ApJ...908...75B}.} We adopt the widely-used method to calculate the time evolution of radio luminosity, where we parametrize the efficiencies of electron acceleration and magnetic field amplification and describe the spectral energy distribution of electrons taking synchrotron cooling and inverse Compton cooling effect into account \citep{2006ApJ...641.1029C, 2006ApJ...651..381C}. Here we employ the bolometric luminosity light curve model of SN\,2004et as a benchmark of type II SNe \citep{2006MNRAS.372.1315S}. The radio luminosity is calculated by considering both synchrotron self-absorption and free-free absorption. For simplicity, we consider a one-zone model to treat the emitting region, in which the nature of the emission is mainly determined by the forward shock properties at the given time. This allows us to clearly visualize the influence of the peculiar time evolution of the forward shock on the radio light curves.\footnote{However, we note that in the actual observational fitting, the multi-zone modeling may be plausible because the regions that have experienced shock compression would be stretched in the radial direction. Then it would be important to take into account the spatial distribution of the electron spectrum and magnetic field in the whole of the shocked CSM.} {For reference we also plot the radio light curves observed in SN~2001ig and SN~2003bg at the frequency of $4.8\,$GHz \citep{2004MNRAS.349.1093R, 2006ApJ...651.1005S}, both of which had been reported to contain time variability during their decay phases.}

In the initial phase within $t\lesssim 10\,{\rm days}$ since the shock breakout from the progenitor, the forward shock is sweeping up the confined CSM characterized by higher densities. Then strong free-free absorption completely damps the radio emission and we cannot expect any bright radio signal in the centimeter wavelength until the forward shock approaches the outer edge of the confined CSM.
At the moment of plunging into the tenuous wind CSM ($t\sim10\,{\rm days}$), the forward shock accelerates instantly by a factor of 2 colliding with CSM by orders of magnitude dilute than confined CSM.
Then the emitting region gets optically thin to both synchrotron self-absorption and free-free absorption, and the system takes a peak of the radio luminosity at $t\gtrsim50\,{\rm days}$. While this peak could be a clue to constrain the nature of CSM around SNe II \citep[e.g.,][]{2025ApJ...978..138I,2025ApJ...985...51N}, radio emission at the frequency of $5\,{\rm GHz}$ is optically thick in the confined CSM. Thus such a low-frequency radio emission is not able to robustly trace the nature of the confined CSM directly. These behaviors of radio light curves are qualitatively consistent with the previous numerical demonstration \citep{2019ApJ...885...41M}.

However, the peculiar time evolution of the forward shock seen in Figure\,\ref{fig:Vshock} can be reflected in the afterward evolution, especially in the two aspects of the light curve. These clues allow even radio emission in the lower frequency to indirectly diagnose the presence of the confined CSM. Firstly, the decline rates of the optically thin emission in the model with the confined CSM are faster than the model without the confined CSM. This is directly attributed to the temporal evolution of the forward shock velocity; as the shock velocity declines quickly, the resultant light curve would be also characterized by the fast temporal evolution.

One of the lessons we can learn is that if we directly fit the fast decline rate of the light curve without introducing confined CSM, we may obtain misleading results. The fast decline of the optically thin emission prefers either smaller $n$ (flatter density profile of the outer layer of the ejecta) or larger $s$ (steeper density profile in the CSM) on the assumption of the single-component of the CSM \citep{2013ApJ...762...14M}. However, our calculations have proven that the same consequence would be derived by assuming the presence of the confined CSM, even without invoking the variation of $n$ and $s$. In terms of this point, there are multiple solutions, both of which can reproduce fast declining radio light curves. When we treat the decline rate of the optically thin emission to constrain the nature of the SN-CSM interaction system, it can be significant to take the effect of the presence of the confined CSM into account.

Another aspect worth discussing is the brightening of the radio luminosity which we can observe at $t\simeq 270\,{\rm days}$ and $t\simeq2900\,{\rm days}$ in the models with $\dot{M}=10^{-5}\Msunperyear$ and $\dot{M}=10^{-4}\Msunperyear$, respectively. This is caused by the reacceleration of the forward shock when the system has restored to the {evolutionary model without the confined CSM.} According to this reacceleration, the radio luminosity suddenly brightens. Afterward, the evolution converges to the model without confined CSM.

{It is noticeable that the timings when we observe the variability of the optically thin radio emission in the model with $\dot{M}=10^{-5}\Msunperyear$ lies in the roughly same timescale as those seen in SN~2001ig, and SN~2003bg. Given that the brightening in our models are seen only once during the decay phase, our scenario may be applied to radio SNe possessing single modulation after the peak as seen in SN~2003bg. It might be possible that this is partially contributing to multiple modulations in SN~2001ig or strong brightening a few years after the explosion as seen in SN~2014C \citep{2017ApJ...835..140M}, while previous studies proposed the origins of the variability and strong brightening as the density variation of the CSM \citep{2004MNRAS.349.1093R,2006ApJ...651.1005S}. Although our models are not tuned to the observational data of SN~2001ig and SN~2003bg and the discussion is limited to qualitative points, we suggest that investigating the inner structure of the CSM could be useful to infer the origin of late-phase radio brightening.}

We suppose that the spiky behavior of this brightening may be caused by the limitation of the one-dimensional configuration in our simulations. Indeed, there should be a Rayleigh-Taylor instability developing and it would induce the mixing process at the interface between different components of gas (i.e., confined CSM - tenuous wind and SN ejecta - confined CSM). This mixing process would alleviate the spiky density structure and the sudden brightening in Figure\,\ref{fig:Lnu}. However, the restoration of the system {to the evolutionary model without confined CSM} is a physically reasonable process proven by our simulations. Hence we expect that the actual behavior of the observed light curve should involve the brightening, but in the more smooth way than illustrated in Figure\,\ref{fig:Lnu}.

\section{Discussion and Summary}\label{sec:summary}

In this study, we have numerically demonstrated the hydrodynamics of the interaction between SN ejecta and CSM, with an emphasis on the evolution after the forward shock has completely swept up the dense confined CSM and plunges into the outside tenuous wind CSM. We prepare three CSM models with varying density scales to examine to investigate the impact of the confined CSM on the afterward evolution. We then compare the simulation results with the evolution of the SN-CSM interaction system without confined CSM. Although our models are presumed to mimic a type II SN progenitor accompanied by confined CSM, we expect that our results can be applied to the other types of SNe such as stripped-envelope SNe, once the appropriate conditions have been satisfied.

In terms of the evolution after the forward shock plunges into the tenuous CSM, we find two important characteristic behaviors observed only in the models with confined CSM. First, we find that the velocity of the forward shock starts deceleration more rapidly at a certain moment of the evolution. The shocked shell consists of the tenuous wind component, and its expansion is driven not by the SN ejecta, but by the ram pressure of the confined CSM component. The fast deceleration of the forward shock begins once the confined CSM component reaches the profile of the homologous expansion, whose typical timescale would be a factor of adiabatic expansion timescale proportional to $M_{\rm CSM}^{1/2}$. As demonstrated in Figure\,\ref{fig:Vshock} and \ref{fig:Lnu}, this deceleration takes place typically $30-300$~days after the explosion and is expected to appear as a rapid decay of optically thin emission in radio light curves.

Second is that the forward shock velocity can again increase by several tens of percentages after the abovementioned fast deceleration. This is caused by the arrival of the reverse shock propagating through the confined CSM component at the SN ejecta. As a result, the ram pressure of the SN ejecta can act on the shocked shell consisting of the tenuous wind, driving the expansion of the shocked shell. This happens when the shocked tenuous wind component accumulates as massive as the total confined CSM mass. This behavior can be reflected as a brightening of the radio luminosity as demonstrated in Figure\,\ref{fig:Lnu}.

{We do not include radiative cooling in our simulations for simplicity, which is potentially important for characterizing the hydrodynamical profiles in terms of the formation of the cooling shell \citep[e.g.,][]{1982ApJ...259..302C,1994ApJ...420..268C}. This cooling shell is believed to be the site where various kinds of radiation in the wavelength from radio, optical, to X-ray are produced so taking into account the formation of this thin shell is likely to be significant for modeling observational signatures. However, it is invariable regardless of the inclusion of radiative cooling that the expansion driving source would change before and after the forward shock finishes sweeping up all of the confined CSM. While we expect that there should be a further alteration of the hydrodynamics profile due to radiative cooling, our results attributed to the change of the driving source of the expansion would be still unchanged.
}

It has been thought that rapid follow-up observations of SNe immediately after the explosion were the only method to investigate the immediate circumstellar environment of SN progenitors. Specifically, the detection of highly ionized flash features and the measurement of early excess in optical light curves within the first few days after the explosion have been highlighted as indicators of the presence of dense confined CSM. However, our study has shown that the effects of the early-phase interaction remain prominent and observable for years after the explosion by demonstrating that the hydrodynamical evolution of the SN-CSM interaction system differs from the self-similar solution classically employed in the literature. This suggests a unique possibility to examine the presence or density of dense confined CSM in the vicinity of the progenitor, even from observations conducted years after the explosion, which should trace the CSM farther from the progenitor.

The presence of the confined CSM could also affect radiative properties in the late phase in other wavelengths. For example, X-ray emission radiated from the forward shock in SNe includes thermal bremsstrahlung emission, nonthermal inverse Compton emission, and possibly synchrotron emission \citep{1996ApJ...461..993F,2006ApJ...641.1029C,2006ApJ...651..381C,2006A&A...449..171N}. Since all of them depend on the temporal evolution of the forward shock velocity, the resultant X-ray light curve could be affected accordingly. The reverse shock propagating through the confined CSM component discussed in Figure\,\ref{fig:hydrodynamics_profile} and \ref{fig:density_mass} heats up the material up to $\gtrsim 10^7\,{\rm K}$. Although its thermal energy density is still subordinate to the ram pressure, this component could be imprinted in X-ray light curves and spectra. We can also consider the possible contributions in optical photometry modeling. In general, optical photometric evolution of SNe without massive CSM is supposed to be characterized by the expanding ejecta and it is sometimes assumed that the presence of the confined CSM could be ignored in the late phase after it is swept up by the forward shock \citep[i.e,. a few tens of day after the explosion,][]{2022A&A...660A..41M}. However, our calculations raise a concern that when fitting the late-phase photometric data without taking into account the presence of initial CSM interaction history might cause misleading interpretations. We believe that it could be more reliable to conduct consistent calculations throughout the early and late phases as demonstrated in \citet{2024A&A...681L..18B,2024A&A...683A.154M}. These considerations would warrant future investigations of SN radiative properties with particular attention to the aftermath of the confined CSM in the late phase.

Peculiar properties seen in the late-phase of SN evolution have been interpreted in terms of physical processes at specific times.
Rapid decline rates of light curves are often fitted using rapidly evolving hydrodynamical models, such as those with small $n$ and large $s$ \citep{2021ApJ...918...34M}. Excess in optical light curves \citep{2020ApJ...900...11W, 2022ApJ...936..111K} and radio rebrightening a few years after the SN explosion \citep{2017ApJ...835..140M, 2023ApJ...954L..45M, 2023ApJ...945L...3M,2024MNRAS.534.3853R,2025arXiv250201740S} are often explained by an additional CSM component originating from eruptive mass loss of the progenitor and rebrightening of radio emission. Our study does not rule out such classical interpretations but rather suggests the possibility that rapid declines or rebrightening can be naturally expected when considering SNe initially interacting with confined CSM.
The rapid decline starts when the confined CSM component reaches homologous expansion and its initiation timescale could be roughly adiabatic timescale of the confined CSM component in a proportional relation of $\propto M_{\rm CSM}^{1/2}$. The rebrightening happens when the confined CSM component has been swept up by the reverse shock. This timing can be roughly given by the moment when the forward shock sweeps up the tenuous wind as massive as the confined CSM mass budget. Note that these two peculiar behaviors will appear together. These hints indicate that, given the mass of the confined CSM, we can predict the characteristic timescales of the forward shock velocity variations seen in Figure\,\ref{fig:Vshock} and even the associated radiative properties in the subsequent phase {as demonstrated in Figure\,\ref{fig:Lnu}}.
To fully understand the properties of SN-CSM interaction occurring a long time after the explosion, it is crucial to take into account their early-phase SN-CSM interaction properties as well. We emphasize the necessity of constructing an interpretation that is consistent across a wide range of SN ages to better understand the final mass-loss activities of massive stars.

\section*{Acknowledgements}
The authors appreciate the anonymous referee for the fruitful comments that improved our manuscript. T.M. and K.-J.C. are supported by the National Science and Technology Council, Taiwan under grant No. MOST 110-2112-M-001-068-MY3, {113-2112-M-001-028-, and NSTC 114-2112-M-001-012-,} and the Academia Sinica, Taiwan under a career development award under grant No. AS-CDA-111-M04. K.M. is supported by the Japan Society for the Promotion of Science (JSPS) KAKENHI grant Nos. 24KK0070 and 24H01810.


\bibliographystyle{aasjournal}
\bibliography{manuscript}{}

\end{document}